\documentclass[conference]{IEEEtran}
\usepackage[T1]{fontenc}

\IEEEoverridecommandlockouts
\usepackage{textcomp}
\usepackage{stfloats}
\usepackage{url}
\usepackage{verbatim}
\usepackage{graphicx}
\usepackage{svg}
\usepackage{enumitem}
\usepackage{soul}

\usepackage[style=ieee,backend=biber]{biblatex}

\usepackage{amssymb} 
\usepackage[normalem]{ulem}
\usepackage{booktabs}
\usepackage[colorlinks=true,
            linkcolor=black,
            urlcolor=black,
            citecolor=black]{hyperref}
            
\usepackage{cleveref}
\usepackage{multirow}
\usepackage{makecell}
\usepackage{bm}
\usepackage{xcolor}
\usepackage{xcolor}
\usepackage[table]{xcolor}
\definecolor{darkgreen}{RGB}{0, 128, 0}
\definecolor{darkred}{RGB}{139, 0, 0}
\definecolor{delftblue}{RGB}{0,166,214}
\definecolor{custombeige}{HTML}{D0CCBE}

\definecolor{M3darkgreen}{HTML}{8DCB20}
\definecolor{M3midgreen}{HTML}{89A655}
\definecolor{M3lightgreen}{HTML}{BAF343} 

\definecolor{delft}{HTML}{00B0F0}

\makeatletter
\def\ps@IEEEtitlepagestyle{%
  \def\@oddhead{\hbox{}\scriptsize 2026 22nd International Conference on the European Energy Market (EEM)\hfill}%
  \def\@evenhead{\hbox{}\scriptsize 2026 22nd International Conference on the European Energy Market (EEM)\hfill}%
  \def\@oddfoot{}%
  \def\@evenfoot{}%
}
\makeatother

\begin{document}

\title{Deep Learning for Electricity Price Forecasting: \\ 
A Review of Day-Ahead, Intraday, and  Balancing Electricity Markets}

\author{Runyao Yu$^{1,2}$, Derek W. Bunn$^3$, Julia Lin$^2$, Jochen Stiasny$^1$, Fabian Leimgruber$^2$, Tara Esterl$^2$,  \\  Yuchen Tao$^4$, Lianlian Qi$^{2,5}$,  Yujie Chen$^6$, Wentao Wang$^7$,  Jochen L. Cremer$^{1,2}$\\ $^1$Delft University of Technology, $^2$Austrian Institute of Technology, $^3$London Business School, $^4$RWTH Aachen \\   $^5$Technical University of Munich, $^6$The Chinese University of Hongkong, $^7$University of Technology Sydney}

\maketitle

\begin{abstract}
Electricity price forecasting (EPF) plays a critical role in power system operation and market decision-making. While prior review studies have summarized forecasting horizons, market mechanisms, and evaluation practices, the rapid adoption of deep learning has introduced a growing diversity of model architectures, output structures, and training objectives that is difficult to compare consistently across studies. 
This paper presents a structured review of deep learning methods for EPF in day-ahead, intraday, and balancing markets. 
We introduce a unified taxonomy that decomposes deep learning models into backbone, head, and loss components. Using this framework, we characterize how modeling choices differ across markets and how they have evolved over time.
Our study highlights the shift toward probabilistic, microstructure-centric, and market-aware designs. We further identify key gaps in the literature, including limited attention to intraday and balancing markets and the need for market-specific modeling strategies, thereby helping to consolidate and advance existing review studies.
\end{abstract}


\begin{IEEEkeywords}
Electricity Price Forecasting, Day-Ahead Market, Intraday Market, Balancing Market, Literature Review.
\end{IEEEkeywords}

\section{Introduction}
\label{introduction}

Europe has one of the world’s most active and tightly coupled wholesale electricity markets, where cross-border trading makes prices a central signal for system operation. In this setting, the day-ahead market provides the primary reference prices through a centralized auction clearing all delivery hours for the next day; however, forecast errors, renewable intermittency, and demand uncertainty leave substantial risk close to delivery~\cite{11050275}. To manage these deviations, European markets rely on continuous intraday trading, which enables participants to adjust positions nearer to real time~\cite{NICKELSEN2025124975}. Remaining mismatches between schedules and physical reality are finally resolved in the balancing market, where Transmission System Operators (TSOs) activate reserves and settle real-time imbalances~\cite{VANDERVEEN2016186}. 
Other regions (e.g., the US) operate real-time markets that are analogous to Europe’s near-real-time layers, but differ in design and are not considered here.
Accordingly, effective trading participation in each market stage relies on accurate Electricity Price Forecasting (EPF). EPF is therefore critical across the full market stack, supporting bidding strategies, congestion handling, and renewable integration.
\begin{figure}[!ht]
  \hspace*{-2.23mm}
\includegraphics[width=1.03\linewidth]{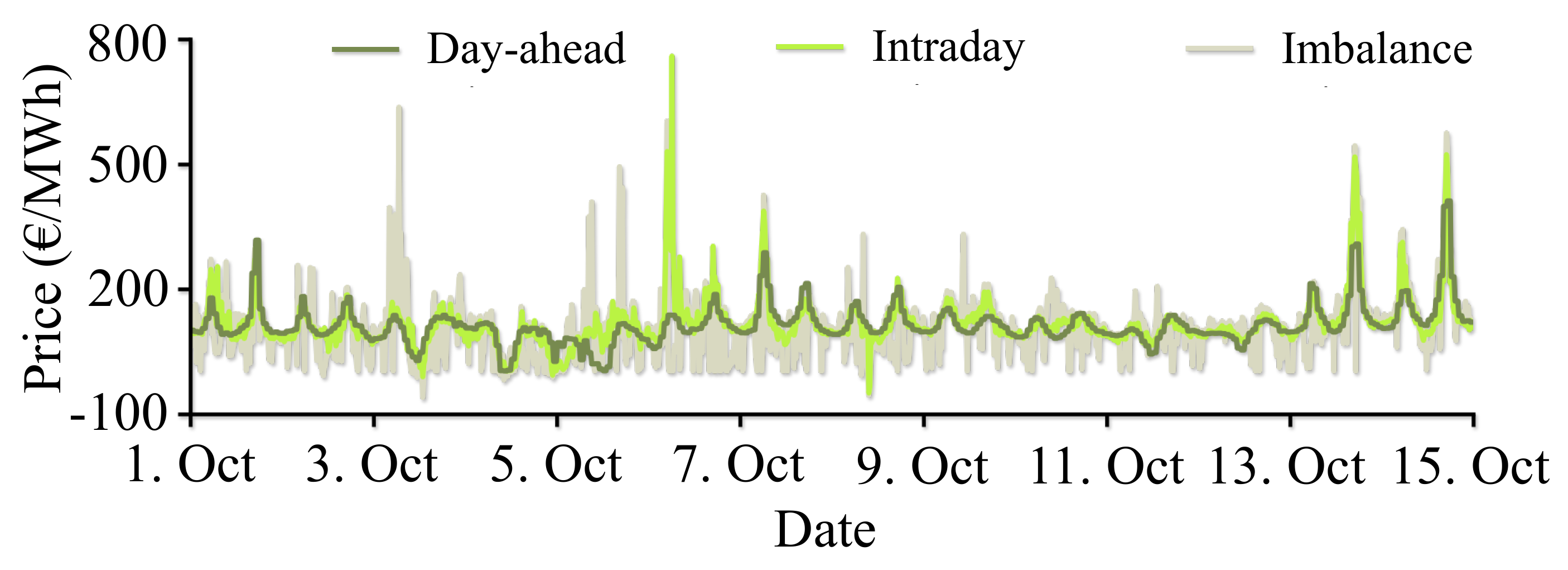}
\caption{
Visualization of day-ahead, intraday, and imbalance prices from 1-15 October 2025 in Austria.
The imbalance price exhibits the highest volatility, the day-ahead price is the smoothest, and the intraday price lies in between.
}
    \label{fig:buy-sell*}
\end{figure} 

Yet EPF is challenging because electricity prices are shaped by non-storability (despite rapid growth in batteries), network constraints, and weather-driven uncertainty, leading to pronounced nonlinearity, regime changes, and heavy-tailed spikes. These difficulties manifest differently across markets: 
day-ahead prices reflect system-wide fundamentals and expectations at hourly (increasingly quarter-hourly) granularity; intraday prices are additionally driven by micro-orderbook dynamics and rapidly updating forecasts; and balancing prices are tightly linked to system imbalance and activation rules, often yielding the highest volatility, as illustrated in Fig.~\ref{fig:buy-sell*}. Consequently, the highly dynamic and spike-prone nature of electricity prices calls for forecasting models that can capture complex dependencies.

\begin{figure*}[!t]
    \centering
    \includegraphics[width=1\textwidth]{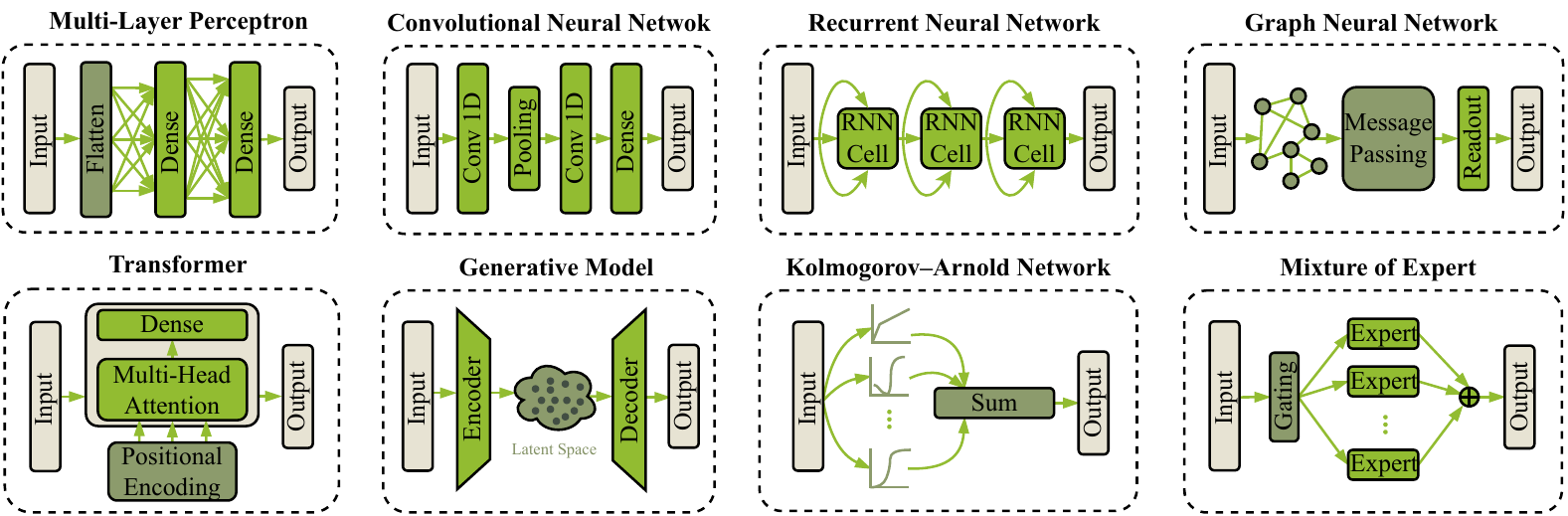}
    \caption{Backbone comparison of deep learning models.}
\label{fig:DLs}
\end{figure*}

\begin{table*}[t]
\caption{Summary of electricity price forecasting studies. }
\label{table_ablation}
\footnotesize
\centering
\begin{tabular}
{
    >{\centering\arraybackslash}p{0.6cm}|
    >{\centering\arraybackslash}p{0.6cm}|
    >{\arraybackslash}p{1.3cm}|
    >{\arraybackslash}p{4cm}|
    >{\arraybackslash}p{0.9cm}|
    >{\arraybackslash}p{1cm}|
    >{\centering\arraybackslash}p{0.8cm}|
    >{\arraybackslash}p{4cm}
}

\toprule
\textbf{Year} & 
\textbf{Paper} & 
\textbf{Market} & 
\textbf{Backbone} & 
\textbf{Head} & 
\textbf{Loss}  & 
\textbf{Feature} & 
\textbf{Country}  \\

\midrule
2018 &\cite{dl1} & Day-ahead & Hybrid (LSTM-MLP / GRU-MLP)  & MMP & MAE & M & Belgium \\
     &\cite{dl3}  & Day-ahead & MLP / CNN / LSTM / GRU  & SMP & MAE &  M & Turkey \\
     &\cite{dl4}  & Day-ahead & MLP & SMP & n.s. & U & Lithuania \\
     &\cite{dl5} & Day-ahead & MLP  & SMP & MSE &  M & Spain / Portugal \\

\midrule
2019

     & \cite{dl16} & Day-ahead & MLP  & SMP & MAE & M & Denmark \\

     & \cite{dl79} & Day-ahead & MLP & SMP & MSE & M & Norway / Sweden / Denmark / +2 \\

    & \cite{E6} & Day-ahead & LSTM  & SMP & n.s. & U & Sweden \\

     & \cellcolor{M3lightgreen}{\cite{dl17}} & \cellcolor{M3lightgreen}{Intraday} & \cellcolor{M3lightgreen}{LSTM / GRU / MLP}  & \cellcolor{M3lightgreen}{SMP} & \cellcolor{M3lightgreen}{MAE} & \cellcolor{M3lightgreen}{M} & \cellcolor{M3lightgreen}{Turkey} \\

      & \cellcolor{M3midgreen}{\cite{E17}} & \cellcolor{M3midgreen}{Balancing} & \cellcolor{M3midgreen}{MLP}  & \cellcolor{M3midgreen}{SMP} & \cellcolor{M3midgreen}{n.s.} & \cellcolor{M3midgreen}{M} & \cellcolor{M3midgreen}{Belgium} \\

\midrule
2020 & \cite{dl2} & Day-ahead & MLP   & SMP & MSE &  M & Germany / Austria \\
     
     & \cite{derek2} & Day-ahead & MLP  & SMP & MSE & M & Spain / Portugal \\
     
     & \cite{dl74} & Day-ahead & MLP  & SMP & n.s. & M & Belgium / France / Germany / + 4 \\

     & \cite{dl78} & Day-ahead & MLP  & SMM & Pinball & M & Norway / Sweden / Denmark / +2 \\

\midrule
2021 & \cite{dl19} & Day-ahead & LSTM / Hybrid (CNN-LSTM)  & SMP & MSE & M & Norway / Sweden / Denmark / + 4 \\

     & \cite{dl20} & Day-ahead & MLP  & SMP & MAE & M & Belgium / Norway / France / + 4  \\

     & \cite{dl80} & Day-ahead & MLP & SMP & MSE & M & Italy \\

\midrule

 2022 & \cite{dl22} & Day-ahead & CNN / MLP  & MMP & LogCosH & M & France / Germany / Belgium \\
 & \cite{dl23} & Day-ahead & LSTM  & SMP & n.s. & M & Sweden \\
 & \cite{dl24} & Day-ahead & Hybrid (Attention-LSTM)  & SMP & MAE & M & Denmark \\
 & \cite{dl27} & Day-ahead & MLP  & SMP & MAE & M & Spain \\

 & \cite{E15} & Day-ahead & MLP / LSTM  & SMP & MSE & M & Norway \\

 & \cellcolor{M3midgreen}{\cite{dl69}} & \cellcolor{M3midgreen}{Balancing} & \cellcolor{M3midgreen}{Hybrid (LSTM-Attention)}  & \cellcolor{M3midgreen}{SMM} & \cellcolor{M3midgreen}{Pinball} & \cellcolor{M3midgreen}{M} & \cellcolor{M3midgreen}{Belgium} \\

 & \cellcolor{M3midgreen}{\cite{dl21}} & \cellcolor{M3midgreen}{Balancing} & \cellcolor{M3midgreen}{MLP}  & \cellcolor{M3midgreen}{SSM} & \cellcolor{M3midgreen}{LogLik} & \cellcolor{M3midgreen}{M} & \cellcolor{M3midgreen}{Germany} \\

\midrule
2023

& \cite{dl34} & Day-ahead & MLP  & SMM & NLL & M & Germany \\

& \cite{dl36} & Day-ahead & MLP  & SMP & MAE & M & Belgium / France / Germany / + 5  \\

& \cite{dl37} & Day-ahead & MLP  & MMP & MAE & M & Belgium / France / Germany / + 5  \\

& \cite{dl77} & Day-ahead & MLP & SMM & Pinball & M & Germany \\

 & \cellcolor{M3lightgreen}{\cite{dl75}} & \cellcolor{M3lightgreen}{Intraday} & \cellcolor{M3lightgreen}{MLP}  & \cellcolor{M3lightgreen}{SMM} & \cellcolor{M3lightgreen}{NLL} & \cellcolor{M3lightgreen}{M} & \cellcolor{M3lightgreen}{Germany} \\
 
 & \cellcolor{M3midgreen}{\cite{dl67}} & \cellcolor{M3midgreen}{Balancing} & \cellcolor{M3midgreen}{MLP / LSTM / GRU / Transformer}  & \cellcolor{M3midgreen}{SSM} & \cellcolor{M3midgreen}{Pinball} & \cellcolor{M3midgreen}{M} & \cellcolor{M3midgreen}{UK} \\

\midrule
2024

& \cite{dl52} & Day-ahead & Transformer  & SMM & Pinball & M & Norway / Finland / Poland / +2 \\
 & \cite{dl39} & Day-ahead & GNN  & MMP & n.s. & M & Norway / Sweden / Denmark / + 4  \\

 
 & \cite{dl47} & Day-ahead & MLP / CNN / Transformer / GRU & SMM & LogLik & M & Germany / France / Belgium / + 4  \\

& \cite{dl50} & Day-ahead & GNN  & MMP & MAE & M & Germany / Hungary / Italy / + 15  \\

& \cite{dl56} & Day-ahead & Hybrid (CNN-GRU) / Transformer  & SMP & MSE & M & Germany / Luxembourg \\

& \cite{dl58} & Day-ahead & Hybrid (CNN-LSTM)  & SMP & MSE & U & UK / Germany \\

 & \cellcolor{M3lightgreen}{\cite{dl40}} & \cellcolor{M3lightgreen}{Intraday} & \cellcolor{M3lightgreen}{LSTM / CNN}  & \cellcolor{M3lightgreen}{SMP} & \cellcolor{M3lightgreen}{n.s.} & \cellcolor{M3lightgreen}{M} & \cellcolor{M3lightgreen}{Denmark} \\
 
 & \cellcolor{M3midgreen}{\cite{dl48}} & \cellcolor{M3midgreen}{Balancing} & \cellcolor{M3midgreen}{MLP / Hybrid (LSTM-MLP)}  & \cellcolor{M3midgreen}{SMP} & \cellcolor{M3midgreen}{n.s.} & \cellcolor{M3midgreen}{M} & \cellcolor{M3midgreen}{Ireland} \\
 
& \cellcolor{M3midgreen}{\cite{dl51}} & \cellcolor{M3midgreen}{Balancing} & \cellcolor{M3midgreen}{Hybrid (LSTM-Attention)}  & \cellcolor{M3midgreen}{SMP} & \cellcolor{M3midgreen}{MSE} & \cellcolor{M3midgreen}{M} & \cellcolor{M3midgreen}{UK} \\

\midrule

2025 & \cite{dl62} & Day-ahead & MLP / LSTM  & SMM & NLL & M & Germany \\

& \cite{dl65} & Day-ahead & RNN  & SMP & MSE & M & France \\

   & \cellcolor{M3lightgreen}{\cite{dl76}} & \cellcolor{M3lightgreen}{Intraday} & \cellcolor{M3lightgreen}{GM}  & \cellcolor{M3lightgreen}{SMM} & \cellcolor{M3lightgreen}{Custom} & \cellcolor{M3lightgreen}{M} & \cellcolor{M3lightgreen}{Germany} \\

& \cellcolor{M3midgreen}{\cite{dl68}} & \cellcolor{M3midgreen}{Balancing} & \cellcolor{M3midgreen}{LSTM}  & \cellcolor{M3midgreen}{SMP} & \cellcolor{M3midgreen}{n.s.} & \cellcolor{M3midgreen}{M} & \cellcolor{M3midgreen}{Greece} \\

& \cellcolor{M3midgreen}{\cite{dl63}} & \cellcolor{M3midgreen}{Balancing} & \cellcolor{M3midgreen}{MLP / Hybrid (LSTM-MLP)}  & \cellcolor{M3midgreen}{SMM} & \cellcolor{M3midgreen}{Pinball} & \cellcolor{M3midgreen}{M} & \cellcolor{M3midgreen}{Ireland} \\

\midrule

2026 & \cite{dl73} & Day-ahead & RNN  & SMP & MSE & M & Germany \\
 
& \cite{dl72} & Day-ahead & MLP  & SMP & MAE & M & Germany / Spain \\

 & \cite{dl0} & Day-ahead & Hybrid (MoE-GNN)  & MMM & Pinball & M & Germany / Austria / France / + 21 \\

  & \cellcolor{M3lightgreen}{\cite{yu2025orderbookfeaturelearningasymmetric}} & \cellcolor{M3lightgreen}{Intraday} & \cellcolor{M3lightgreen}{KAN / MLP}  & \cellcolor{M3lightgreen}{SSM} & \cellcolor{M3lightgreen}{Pinball} & \cellcolor{M3lightgreen}{M} & \cellcolor{M3lightgreen}{Germany / Austria} \\

 & \cellcolor{M3lightgreen}{\cite{dl71}} & \cellcolor{M3lightgreen}{Intraday} & \cellcolor{M3lightgreen}{Hybrid (MLP-Cross-Attention)}  & \cellcolor{M3lightgreen}{SSM} & \cellcolor{M3lightgreen}{Pinball} & \cellcolor{M3lightgreen}{M} & \cellcolor{M3lightgreen}{Germany / Austria} \\

 & \cellcolor{M3midgreen}{\cite{yu2026marketruleinformedneuralnetworkefficient}} & \cellcolor{M3midgreen}{Balancing} & \cellcolor{M3midgreen}{Hybrid (Prior-Guided MLP)}  & \cellcolor{M3midgreen}{SSM} & \cellcolor{M3midgreen}{Pinball} & \cellcolor{M3midgreen}{M} & \cellcolor{M3midgreen}{Austria} \\
\bottomrule
\end{tabular}
\end{table*}


Against this backdrop, deep learning has emerged as a powerful tool for EPF by learning complex nonlinear patterns from data, with common architectures including
Multi-Layer Perceptron (MLP),
Long Short-Term Memory (LSTM), and Convolutional Neural Network (CNN)~\cite{dl1,dl3,dl4,dl5}. Subsequent work expanded the scope to multi-timestep forecasting, multi-market settings, and probabilistic outputs~\cite{dl16,dl78,dl19,dl20}. More recently, Transformers, graph-based models, and Mixture-of-Experts (MoEs) have further enriched the modeling landscape, enabling long-range dependency modeling, spatial coupling across bidding zones, and scalable multi-country forecasting~\cite{dl24,dl39,dl50,dl52,dl0}.

Existing review papers have established a strong foundation for EPF by clarifying market mechanisms and forecasting horizons~\cite{review6,review1,review7}. Other studies have emphasized the importance of rigorous and reproducible evaluation, advocating standardized datasets, error measures, and benchmarking toolkits~\cite{review4,review3}. In parallel, probabilistic EPF has been systematically formalized through reliability-sharpness principles and statistically sound evaluation protocols~\cite{review5}. More recently, cross-market surveys have begun to jointly consider day-ahead, intraday, and balancing markets for general method analysis~\cite{review2}. 
Nevertheless, as deep learning plays an increasingly central role in EPF, most existing surveys provide limited in-depth analysis of {deep learning}. In particular, the interplay between representation learning, output formulation, and optimization objectives, and the way these design choices encode market characteristics, remains insufficiently structured. This gap motivates a unified deep learning taxonomy that can explain methodological shifts in the literature and enable consistent synthesis across markets.

To organize this rapidly growing and heterogeneous literature, this paper adopts a unified modeling perspective that decomposes deep learning approaches into three fundamental components: 
 {backbone}, {head}, and {loss function}.
Using this framework, we review and conclude trends in deep learning-based EPF across three markets. 
The main contributions of this paper are as follows:
\begin{itemize}
    \item \textbf{Unified deep learning evaluation taxonomy:} 
    We propose a unified evaluation framework that decomposes deep learning models into backbone, head, and loss components, and provides a consistent evaluation perspective for comparing architectures, output structures, and forecasting objectives across studies.

    \item \textbf{Cross-market trend synthesis:} Using this taxonomy, we provide a comprehensive synthesis of deep learning trends in day-ahead, intraday, and balancing markets, clarifying how modeling choices evolve in response to distinct market characteristics and where similar design patterns recur across markets.
    
     \item \textbf{Identification of gaps and future directions:} 
     We highlight key gaps and open challenges in the literature, including limited attention to intraday and balancing markets, the increasing importance of microstructure-aware and trajectory-based forecasting, and the need for market-specific design in deep learning models.

\end{itemize}


The remainder of this paper is organized as follows. 
Section~\ref{model} presents the fundamental components of deep learning models, including the backbone, head, and loss function. Section~\ref{trends} analyzes trends and discusses future research directions in deep learning for EPF across markets. Section~\ref{sec:conclusion} concludes the paper and summarizes key findings.


\section{Deep Learning Components}
\label{model}

A deep learning model typically consists of three components: a \textbf{backbone}, a \textbf{head}, and a \textbf{loss function}. 
The \emph{backbone} defines how a deep learning model processes the input sequence \(\mathbf{X} \in \mathbb{R}^{L \times F}\), where \(L\) is the number of timesteps and \(F\) the number of features (e.g., load, wind, price). 
The \emph{head} varies depending on whether the model is trained for single- or multi-output forecasting. The \emph{loss} may be pointwise (e.g., MAE) or probabilistic (e.g., quantile loss).

\subsection{Backbone}
\label{sec:backbones}

The choice of backbone reflects assumptions about temporal, spatial, and structural patterns in electricity markets.
The simplest backbone is the \textbf{MLP}, which flattens the input sequence into a vector of size \(L \cdot F\). While computationally efficient, MLPs ignore sequential or spatial dependencies.
To retain temporal structure, \textbf{CNNs} operate directly on input sequences, applying local convolutional filters over time to capture short-range patterns. 
\textbf{Recurrent Neural Networks (RNNs)} also take a sequential view, processing input sequence step-by-step to build memory over time. Variants like \textbf{LSTM} and \textbf{GRU} mitigate vanishing gradients for long-sequence modeling, but are slow to train due to limited parallelism.
Replacing recurrence with an \textbf{Attention} mechanism, \textbf{Transformer} processes the entire sequence in parallel, learning dependencies between all time steps. 
Extensions such as \textbf{Cross-Attention} further allow Transformers to condition price forecasts on contextual information.
As electricity markets are spatially interconnected through transmission lines, \textbf{Graph Neural Networks (GNNs)} extend temporal backbones by explicitly incorporating graph structures of transmission topology, where the spatial dependencies are captured via message passing over the graph.
\textbf{Generative Models (GMs)} such as \textbf{Variational Autoencoder (VAE)} and \textbf{Generative Adversarial Network (GAN)}, 
learn distributions over input sequence, offering calibrated probabilistic forecasts.
A recent addition is the \textbf{Kolmogorov–Arnold Network (KAN)}, which expresses functions over input sequences via sparse sums of univariate nonlinearities, yielding interpretable and structured representations.
In scenarios with high data heterogeneity, \textbf{MoE} improve flexibility by routing input sequences through multiple specialized subnetworks (e.g., MLP or CNN), with a learnable gating function selecting which experts to activate. This conditional computation enables the model to adapt to different regimes and load patterns.
When one backbone is not enough, 
\textbf{Hybrid Models} combine different modeling paradigms: CNNs can extract local features before feeding them into LSTMs~\cite{qin2017dual}, attention mechanisms can replace MoE-style gating for dynamic expert selection~\cite{li2021enhancing}, and domain-specific prior knowledge, such as imbalance-pricing rules, can be embedded into neural networks to improve parameter efficiency~\cite{yu2026marketruleinformedneuralnetworkefficient}.
The architectural choices reflect different assumptions about how electricity price signals evolve across time and space.

\subsection{Head}
\label{sec:heads}

The \emph{head} determines the output structure of the model, reflecting how many countries, timesteps, and quantiles are forecasted simultaneously. Given the encoded representation of the input \(\mathbf{X} \in \mathbb{R}^{L \times F}\), the head maps it to a structured output depending on the forecasting objective.
The output structure can be summarized by a three-letter code. The first letter indicates whether the model forecasts a single or multiple countries (\textbf{S} or \textbf{M}), the second indicates a single or multiple timesteps (\textbf{S} or \textbf{M}), and the third indicates whether the output is pointwise, a single quantile, or multiple quantiles (\textbf{P}, \textbf{S}, or \textbf{M}). 
For example, \textbf{SSP} denotes single-country, single-timestep point forecasting, whereas \textbf{MMM} denotes multi-country, multi-timestep, multi-quantile forecasting.


Beyond standard formulations, the flexibility of neural networks also enables more advanced multi-head designs with additional structural constraints. For instance, \cite{dl71, dl0} proposed a \emph{hierarchical quantile head} in which the median quantile is first predicted using a dense layer, and  a non-negative residual is learned through another dense layer. Then, the upper quantile prediction is constructed by adding the non-negative residual to the median prediction, which effectively mitigates the quantile crossing issue, as also discussed in prior works~\cite{sortquantilecrossing1, sortquantilecrossing2, dl75}.

\subsection{Loss Function}
\label{sec:loss}

The choice of loss function depends on the forecasting objective, which may be pointwise, probabilistic, or customized to reflect domain-specific considerations. 

For {pointwise forecasting}, common loss functions include \textbf{Mean Squared Error (MSE)} and \textbf{Mean Absolute Error (MAE)}, defined as:
\begin{equation}
    \mathcal{L}_{\mathrm{MSE}} = \frac{1}{N} \sum_{i=1}^N (y_i - \hat{y}_i)^2,
\end{equation}
\begin{equation}
    \mathcal{L}_{\mathrm{MAE}} = \frac{1}{N} \sum_{i=1}^N |y_i - \hat{y}_i|,
\end{equation}
where \(N\) denotes the number of samples, and \(y_i\) and \(\hat{y}_i\) denote the true and predicted values. MSE penalizes large errors more strongly due to its quadratic form, making it more sensitive to outliers compared to MAE.

For {probabilistic forecasting}, two major classes of loss functions are commonly used. The first is the \textbf{Pinball Loss}, widely used for training quantile regressors. For a given quantile level \(\tau \in (0,1)\), the pinball loss is defined as:
\begin{equation}
    \mathcal{L}_{\mathrm{pinball}}(\tau) = \frac{1}{N} \sum_{i=1}^N 
    \max\left( \tau (y_i - \hat{y}_i), (\tau - 1)(y_i - \hat{y}_i) \right).
\end{equation}
This formulation encourages correct quantile calibration by penalizing overestimation and underestimation differently. The second is the \textbf{Log-Likelihood (LogLik)}. Assuming a parametric form (e.g., Gaussian), the LogLik is maximized, or \textbf{Negative Log-Likelihood (NLL)} is minimized, when the model assigns high probability density to the observed targets:
\begin{equation}
    \mathcal{L}_{\mathrm{LogLik}} =  \frac{1}{N} \sum_{i=1}^N \log p(y_i \mid \theta_i),
\end{equation}
where \(p(y_i \mid \theta_i)\) is the predicted likelihood of the true value \(y_i\) under distribution parameters \(\theta_i\) (e.g., predicted mean and variance). LogLik-based losses are commonly used in generative models and parametric uncertainty quantification.

In addition to standard objectives, one may design {customized loss functions} to guide the model toward desired behavior. For example, \cite{maciejowska2025statisticaleconomicevaluationforecasts} proposes a Min-Max Price Deviation (MPD) loss, which reflects the profitability of battery systems. 

\section{Trends in Electricity Price Forecasting}
\label{trends}
This section reviews empirical trends in deep learning-based EPF across day-ahead, intraday, and balancing markets. Despite differences in market design and forecasting objectives, several common patterns emerge. As shown in Table~\ref{table_ablation} (n.s. denotes \emph{not specified}; U and M under the \emph{Feature} column indicate univariate and multivariate inputs), most studies favor {multivariate} input features, combining historical prices with exogenous variables such as load and renewable generation, reflecting the strong fundamental drivers of price dynamics. At the same time, a portion of the literature does not explicitly specify the training loss, complicating direct comparison across models. Beyond these shared characteristics, modeling choices diverge substantially by market, as discussed in the following subsections.

\subsection{Day-Ahead Market}
\label{sec:da_review}

Day-ahead price forecasting has been the earliest and most extensively studied application of deep learning in power markets. As summarized in Table~\ref{table_ablation}, the literature from 2018 to 2026 reveals several clear methodological trends.

\paragraph{Evolution of model families} Early studies (2018-2020) predominantly relied on simple backbones such as MLP, LSTM, GRU, and CNN, either individually or in hybrid forms~\cite{dl1,dl3,dl4,dl5,dl16}. From 2021 onward, more expressive architectures began to emerge: attention-based models were increasingly adopted to capture temporal dependencies~\cite{dl24,dl52}, while GNNs were introduced to model spatial dependencies across interconnected bidding zones~\cite{dl39,dl50}. 
Most recently, large-scale architectures combining MoE and graph structures have been proposed to jointly model multiple regions and market conditions~\cite{dl0}, where graph priors encode physical and market connectivity across bidding zones, marking a shift toward foundation-style models.

\paragraph{Preference for multi-timestep heads} Because day-ahead markets require forecasts for all delivery times of the next day, most deep learning models adopt {multi-head designs}. As shown in Table~\ref{table_ablation}, the dominant head type is {SMP} (single-country, multi-timestep, point forecasting)~\cite{dl3,dl4,dl5,dl16,dl19}. In more advanced settings, models extend this formulation to {MMP}, {SMM}, or MMM heads, enabling simultaneous forecasting across multiple countries and/or multiple quantiles~\cite{dl22,dl36,dl37,dl0}. 

\paragraph{Shift from pointwise to probabilistic setting} Another notable trend is the gradual transition from pointwise objectives toward probabilistic forecasting. Prior to 2022, most studies optimized pointwise losses such as MAE or MSE~\cite{dl1,dl3,dl16,dl19}. After 2023, however, an number of works adopted probabilistic losses to better capture uncertainty in day-ahead prices~\cite{dl34,dl52,dl47,dl78}. This shift reflects growing recognition that point forecasts alone are insufficient for risk-aware decision making, especially under high renewable penetration and volatile market conditions.

\subsection{Intraday Market}
\label{sec:id_review}

Compared to day-ahead forecasting, intraday price forecasting has received substantially less attention in the literature. Existing studies nevertheless reveal several emerging trends.

\paragraph{From macro-level features to microstructure}
Early intraday studies typically adopt a macro-level perspective, using aggregated price indices or engineered features, such as Volume-Weighted Average Prices (VWAPs), together with exogenous variables like load and renewable generation. These features are then processed by relatively simple backbones such as MLPs or LSTMs~\cite{dl17,dl40}. More recently, the focus has shifted toward orderbook-centric modeling. Instead of relying on VWAPs,
\cite{yu2025orderbookfeaturelearningasymmetric} extracts 384 orderbook features and demonstrates that price percentiles carry strong predictive power. 
Going one step further, \cite{dl71} directly models the raw orderbook using a cross-attention mechanism that reflects market-specific priors on buy-sell interactions, capturing the dynamics between buy and sell orders and achieving state-of-the-art performance. These works suggest that the predictive potential of orderbook remains largely underexplored.

\paragraph{From single-index prediction to trajectory forecasting}
While early studies focus on one-step-ahead single-index forecasting (e.g., ID\(_3\))~\cite{dl17,dl40}, recent works increasingly emphasize trajectory forecasting~\cite{dl75,hirsch2024simulation}. Predicting an entire price path provides \emph{more informative} signals about the temporal evolution of intraday prices, which can support decision-making beyond isolated point forecasts. Generative and distributional models are particularly well suited for this setting, as they naturally capture temporal uncertainty and price path variability.

\paragraph{Limited accessibility and synthetic data}
Overall, intraday deep learning research remains sparse, with most existing studies concentrating on Germany and Austria~\cite{dl75, yu2025orderbookfeaturelearningasymmetric, dl71}. A key barrier is the limited accessibility of raw orderbook. Moreover, although not studied in a deep learning context, several works have shown that using information from neighboring delivery hours can further improve forecasting performance~\cite{hirsch2024multivariate, hornek2025directional}. 
In this context, synthetic data generation may offer a promising future research direction, for example, by simulating realistic orderbook dynamics and neighboring trade interactions across delivery hours.

\subsection{Balancing Market}
\label{sec:bm_review}

Compared to day-ahead and intraday markets, imbalance prices exhibit stronger volatility, heavy-tailed distributions, and pronounced regime switching driven by market rules and system conditions. As a result, deep learning applications in balancing markets have developed along distinct methodological directions, as summarized below.

\paragraph{Two-stage modeling paradigms}
As imbalance price formation is governed by explicit market mechanisms, 
many studies adopt two-stage modeling paradigms, in which deep learning models are first used to predict intermediate variables (e.g., system imbalance), followed by a dedicated mapping to imbalance prices~\cite{E17,dl21,dl63}. 
This mechanism-aware modeling philosophy distinguishes balancing-market forecasting from more direct end-to-end approaches commonly used in day-ahead and intraday settings.

\paragraph{Preference for shallow backbones}
In contrast to the increasing architectural complexity observed in day-ahead and intraday forecasting, balancing-market studies predominantly rely on shallow backbones with relatively few layers. These lightweight architectures are favored for their interpretability and computational efficiency, particularly when models are deployed in operational decision-making or trading contexts~\cite{E17,dl48,dl63,dl68, yu2026marketruleinformedneuralnetworkefficient}. 

\paragraph{Single-country focus}
Another prominent characteristic is the strong focus on single-country market. Balancing mechanisms may vary across regions in terms of pricing rules, reserve products, and settlement procedures, which has led most studies to concentrate on individual national markets such as Germany~\cite{dl69}, Austria~\cite{yu2026marketruleinformedneuralnetworkefficient}, UK~\cite{dl51,dl67}, and Ireland~\cite{dl48,dl63}. While this market-specific focus enables close alignment with local price formation rules, it also limits cross-country generalization and highlights the challenges of developing unified balancing-market forecasting models.

\section{Conclusion}
\label{sec:conclusion}

This review surveyed deep learning for electricity price forecasting across day-ahead, intraday, and balancing markets through a unified perspective of backbone, head, and loss design. While day-ahead forecasting has evolved toward  probabilistic, multi-market, and foundation-style models, intraday research remains sparse and is shifting toward microstructure-aware and trajectory-based forecasting. Balancing markets, by contrast, favor price-formation-aware design due to strong regime switching rules. 
Future research would benefit from greater attention to intraday and balancing markets, as well as from more advanced model structures that incorporate domain knowledge and market-specific mechanisms while retaining the flexibility of data-driven learning.


\printbibliography[heading=bibintoc,title=References]

\end{document}